\documentclass[12pt,a4paper]{article}
\usepackage{tikz}
\usetikzlibrary{matrix}
\usepackage{amsmath}
\usepackage{amssymb}
\usepackage{amsfonts}
\usepackage{graphicx}
\usepackage{subfig}
\usepackage{caption}
\usepackage{dcolumn}
\usepackage{bm}
\usepackage{cite}
\usepackage[
colorlinks,linkcolor=blue,citecolor=magenta]{hyperref}
\usepackage[top=3cm,right=2cm,bottom=2.5cm,left=2cm]{geometry}

\begin{document}
	\title{Hamiltonian Structure of Fierz-Pauli Gravitons,\\ Partially Massless Fields \\ and Gauge Symmetry}
	
	\author{Jalali Rezvan\thanks{r.jalali@ph.iut.ac.ir} , Shirzad Ahmad\thanks{shirzad@ipm.ir}\\
	 \textit{Department of Physics, Isfahan University of Technology}}
	\maketitle

\begin{abstract}
We study the constraint structure of  Fierz -Pauli action  in  both flat  and curved space in the framework of Hamiltonian formalism. We observe an abrupt change in the constraint algebra and the characteristics of the constraints when the mass term is turned off.
As is well-known, for de Sitter background with a special tuning of the mass, we will have a gauge symmetry and one  degree of freedom less. We will show how this abrupt change in the behavior of the system is reflected on the properties of the constraint algebra.  We also obtain the generating functional of gauge transformations both  for the massless and partially massless theories.
\end{abstract}

\section{Introduction}
\label{sec:intro}
Fierz and Pauli in their pioneer work in 1939 \cite{FP1} suggested their famous Lagrangian for the free massless spin-2 field. Investigating the equations of motion, they found, in four dimensional space,  five degrees of freedom for the massive theory  and two degrees of freedom for the massless one. In arbitrary $D$  dimensions these are $\frac{1}{2}D\left(D-3\right) $ and  $\frac{1}{2}\left( D+1\right) \left( D-2\right) $  respectively. It can also be seen directly that the Fierz-Pauli Lagrangian  is invariant under the gauge transformations
$h_{\mu\nu} \rightarrow h_{\mu\nu} + \partial_{\mu}\xi_{\nu} + \partial_{\nu}\xi_{\mu}$. This symmetry is no longer established for the massive theory.
It was also observed that the massless spin-2 Lagrangian  may be considered as the linearization of the Hilbert-Einstein action  of general relativity for the perturbation $ h_{\mu\nu} $ around the flat metric.

For almost seven decades, no consistent covariant theory was found  whose linearized version corresponds to the massive Fierz-Pauli model. Since 2010 a flux of papers based on the  works of de Rham-Gabadadze \cite{dRG,dRG2} and Hassan-Rosen \cite{HR1} has been published on the issue of the massive gravity. See also refs. \cite{HR2,HR3} and the reviews \cite{derham, Hinterbichler1}. 

As is well known, the Hamiltonian analysis provides a powerful tool for investigating the dynamical properties, including the number of degrees of freedom as well as the gauge symmetries, of a theory (see Refs. \cite{Dirac,Henneaux,Loran,Rothe}). For the Hilbert-Einstein action of general relativity this may be done in the framework of the famous ADM decomposition \cite{Deser0}
For Fierz-Pauli theory the Hamiltonian analysis has been given briefly in Ref. \cite{Hinterbichler1}. However, the existing investigation relies on adding total time derivative terms to the Lagrangian.  In a recent work \cite{Deser00} Deser has given a canonical analysis of the Fierz-Pauli theory in a first order investigation based on considering the six spacial components $h_{ij}$ as dynamical variables. 

Our first task in this paper is a detailed Hamiltonian analysis for the massless and massive Fierz-Pauli theories. We show that the original Lagrangian (with no need to add anything) provides enough constraints to give the correct number of degrees of freedom and generate the symmetry transformation. We have considered all of the covariant components $h_{\mu \nu}$ as dynamical variables. As we will see, this approach enables us to better understand the gauge symmetry of the model in the framework of Hamiltonian formalism.

An interesting point is  different behavior of the  theory in the limit $M=0$, where $M$ is the graviton mass. In fact, the theory behaves discontinuously for $M =0$, i.e. even for small $M$ we have five degrees of freedom and no guage symmetry, while for $M=0$ the emerged gauge symmetry leaves only two degrees of freedom.  As we will see, it turns out very natural that some Poisson brackets among the constraints vanish exactly at the point $ M =0$ of the parameter space. Hence, the corresponding constraints turn out to be first class only at the point $M=0$.

The Fierz-Pauli theory can also be written in a curved space. For this reason, besides converting partial derivatives $\partial_{\mu}$ into covariant derivatives $\nabla_{\mu}$, it is also needed to add a compensating curvature dependent term to retrieve the gauge symmetry  in the covariant form $h_{\mu\nu} \rightarrow h_{\mu\nu} + \nabla_{\mu}\xi_{\nu} + \nabla_{\nu}\xi_{\mu}$. 

Deser-Walderon  in an interesting paper\cite{PM1}(also see 
\cite{PM2,Deser1,Deser2,Deser3,Deser4,Geo1,Geo2,Deser5,Deser6})and long before that, Higuchi \cite{Higuchi} observed that for de Sitter background, if the graviton mass $M$ and cosmological constant $\Lambda$ are related in $D$ dimensions via the relation $\Lambda = (\frac{D-1}{D-2})M^{2} $, then the massive Fierz-Pauli theory acquires a new gauge symmetry. Hence, one degree of freedom would be lost, e.g. in four dimensions, we will have  four degrees of freedom. This theory is recognized as the partially massless(PM) theory . Since, In PM theories the cosmological constant is tied to the graviton mass, these models have acquired so much interest in order to solve the well-known cosmological constant problem. In other words, one may warrant the smallness of $\Lambda$ by introducing a small graviton mass $M$. 

It is aimed thereafter to develop a covariant theory of gravity whose linear version around a standard background leads to a PM theory (see Refs. \cite{Deser10,Deser11,Hinterbichler2,Hinterbichler6,de Rham,Joung,Apolo,Rosen}.) In the next two following sections we will give our Hamiltonian analysis for the Fierz- Pauli theory in flat and de Sitter backgrounds, respectively. This analysis is more technical due to explicit time dependent terms in the Lagrangian. The final result for the number of degrees of freedom is similar  to the flat space.

Our interesting observation is the abrupt behavior of the system just when the theory touches the Higuchi bound \cite{Deser1}, where the constraint structure of the system has a completely different character. Exactly at the Higuchi bound, one chain of constraints  with four second class elements changes to a first class chain with three elements.

We have also employed the technicalities of Refs. \cite{Shabani,pons} for introducing the generating functional of gauge transformations constructed upon the first class constraints of the system. We have done this to find the gauge generating functional for the gauge symmetry of the original massless Fierz-Pauli theory and the Higuchi gauge symmetry of the partially massless theory.
\section{Hamiltonian structure of Fierz-Pauli action in flat space}

In this section, we briefly review the Hamiltonian structure of  Fierz-Pauli model in  massive and massless sectors. The Lagrangian in $D$ dimensions reads \cite{FP1}. 
\begin{align} \label{a1}
S  = \int d^{D}x \sqrt{-g} \,  \Big(-1/2 \partial_{\lambda}h_{\mu\nu}\partial^{\lambda}h^{\mu\nu}
+\partial_{\lambda}h_{\mu\nu}\partial^{\nu}h^{\mu\lambda}-\partial_{\mu}h\partial^{\nu}h^{\mu\nu}+ 1/2\partial_{\lambda}h\partial^{\lambda}h\Big)\ .
\end{align}
where $ h_{\mu\nu}$, as dynamical variables,  are $\frac{1}{2}D(D+1)$  components of a symmetric rank-2 tensor.
For the massive spin-2 field one should add the mass term
\begin{align} 
-\frac{1}{2} M^{2}\left( h_{\mu\nu}h^{\mu\nu} -h^{2}\right) \label{massterm}
\end{align}
to the Lagrangian (\ref{a1}). The coefficient (-1) between the two quadratic terms in Eq. (\ref{massterm}) is crucial and is known as the Fierz-Pauli tuning.
 Separating time and space components of the fields and derivatives, the Lagrangian (1) is written as 
\begin{align}\label{a2}
\mathcal{L} = &\frac{1}{2}\partial_{0}h_{ij}\partial_{0}h_{ij} - \frac{1}{2}\partial_{0}h_{ii}\partial_{0}h_{kk} -\partial_{0}h_{0k}\partial_{k}h_{00} - 2 \partial_{i}h_{0j}\partial_{0}h_{ij}\nonumber\\
& + \partial_{0}h_{0k}\partial_{k}h_{ii} + \partial_{k}h_{0k}\partial_{0}h_{00} + \partial_{k}h_{0k}\partial_{0}h_{ii}- \partial_{i}h_{00}\partial_{i}h_{kk} \nonumber\\
& + \partial_{k}h_{ij}\partial_{i}h_{kj}
+ \frac{1}{2}\partial_{i}h_{kk}\partial_{i}h_{jj} - \frac{1}{2}\partial_{i}h_{jk}\partial_{i}h_{jk} + \partial_{i}h_{ik}\partial_{k}h_{00}\nonumber\\
& - \partial_{i}h_{ik}\partial_{k}h_{jj} + \partial_{i}h_{0k}\partial_{i}h_{0k} - \partial_{i}h_{0k}\partial_{k}h_{0i} \ .
\end{align}
Using Eq. ( \ref{a2}), the canonical momenta are easily derived as
\begin{align}
\label{a3} &\pi^{00}= -\partial_{k}h_{0k} \ , \\ \label{a4} 
& \pi^{0i}=  \partial_{i}h_{00}+ \partial_{i}h_{kk}\ ,\\
&\pi^{ij}= \partial_{0}h_{ij} + \partial_{k}h_{0k}\delta_{ij} - \partial_{0}h_{kk}\delta_{ij}
- 2 \partial_{i}h_{0j} \ . 
\end{align}
Eqs. (\ref{a3}) and  (\ref{a4}) lead to the following primary constraints

\begin{align}
\phi^{00}= \pi^{00} + \partial_{k}h_{0k} \ ,  \quad   \phi^{0i} = \pi^{0i}  - \partial_{i} h_{00} - \partial_{i}h_{kk} \ .
\end{align}\\
So, the total Hamiltonian reads
\begin{align}
H_{T} = \int d^{D}x \ \Big( \mathcal{H}_{c}+ u_{00}\phi^{00} + u_{0i}\phi^{0i}\Big) \label{H_{T}}\ ,
\end{align}
where $u_{00}$ and $u_{0i}$ are Lagrange multipliers and $\mathcal{H}_{c}$ is the canonical Hamiltonian  density. For the massless case $\bar{\mathcal{H}_{c}}$ is as follows,
\begin{align}
 \bar{\mathcal{H}_{c}}= &\frac{1}{2}\left( \pi^{ij}\right)^{2}  -\frac{1}{2} \frac{1}{D-2} \left(\pi^{kk} \right)^{2} - \frac{1}{D-2} \pi^{ii} \partial_{k}h_{0k}- \frac{1}{2}\frac{D-1}{D-2} \left( \partial_{k}h_{0k}\right) ^{2}  \nonumber \\
& + \pi^{ij}\left(\partial_{i}h_{0j}+ \partial_{j}h_{0i} \right)
   + \frac{1}{2} \left(\partial_{k}h_{ij} \right) \left(\partial_{k}h_{ij} \right)\nonumber \\
 &-\frac{1}{2}{\left(\partial_{i}h_{kk} \right)\left(\partial_{i}h_{jj} \right)}  + \left(\partial_{i}h_{0j}\partial_{i}h_{0j} + \partial_{i}h_{0j}\partial_{j}h_{0i} \right)\nonumber\\
&  - {\left(\partial_{i}h_{kj} \right)\left(\partial_{k}h_{ij} \right)} + {\left(\partial_{i}h_{ik} \right)\left(\partial_{k}h_{jj} \right)}- h_{00} \left(  \nabla^{2}h_{kk} -\partial_{i}\partial_{j}h_{ij}\right)\ .
\end{align}
For the massive theory the canonical Hamiltonian density reads
\begin{align}
\mathcal{H}_{c}= \bar{\mathcal{H}_{c}} + M^{2}\Big(\dfrac{1}{2}h^{2}_{ij}+ \frac{1}{2}h^{2}_{kk}- h_{00} h_{kk} - h^{2}_{0i}
\Big)\ .
\end{align}
Using the fundamental Poisson brackets
\begin{align}
& \left\lbrace h_{00}(x) , \ \pi^{00}(y)\right\rbrace = \delta\left( x-y \right), \hspace{5mm}  \left\lbrace h_{0i}(x) , \ \pi^{0j}(y)\right\rbrace = \delta^{j}_{i}\delta\left( x-y\right) \ , \nonumber \\
&  \left\lbrace h_{ij}(x) ,  \ \pi^{kl}(y)\right\rbrace = \delta^{l}_{i}\delta^{k}_{j}\delta\left( x-y\right)\ ,
\end{align}
consistency of primary constraints, for the massless case, leads to the second level constraint $\bar{\chi}^{\ 00}\equiv \left\lbrace \bar{\phi}^{\ 00}  ,\ \bar{H}_{c} \right\rbrace $ and  $\bar{\chi}^{\ 0i} \equiv \left\lbrace \bar{\phi}^{\ 0i}  \ ,\ \bar{H}_{c} \right\rbrace $ as follows
\begin{align}
\bar{\chi}^{\ 00} = \nabla^{2}h_{kk} - \partial_{i}\partial_{j}h_{ij} \ , \hspace{12mm}  \bar{\chi}^{\ 0i} = 2 \partial_{j}\pi^{ij}  + 2 \nabla^{2}h_{0i}\ .
\end{align}
For the massive theory the second level constraints read
	\begin{align}\chi^{00} =\bar{\chi}^{\ 00} - M^{2}h_{kk} \ , \hspace{12mm} \chi^{ 0i} = \bar{\chi}^{\ 0i} + 2 M^{2}h_{0i}\ .
	\end{align}	
Let us first proceed with the massless model. In this case the second level constraints have strongly vanishing Poisson brackets with the primary constraints and weakly vanishing Poisson brackets with the canonical Hamiltonian as follows 
 \begin{align}\label{massless case}
&\left\lbrace \bar{\chi}^{\ 00}  ,\ \bar{H}_{c} \right\rbrace  = - \frac{1}{2}\partial_{i}\bar{\chi}^{\ 0i} \approx 0 \ , \nonumber \\
&\left\lbrace \bar{\chi}^{\ 0i} ,\ \bar{H}_{c} \right\rbrace  = 0 \ .
\end{align}
Hence, we have no third level constraint for the the massless Fierz-Pauli theory.
We have altogether $ 2 D $ first class constraints in two levels.

The number of dynamical degrees of freedom$(DOF)$ should be found through the master equation \cite{Henneaux}
\begin{align}\label{master equation}
DOF = \dfrac{1}{2}(2N - 2FC -SC)\ .
\end{align}
where $ N $, $ FC $ and $ SC $ represent the number of original variables, first class and second class constraints, respectively. Using this formula,  we have, for the massless Fierz-Pauli theory,  $$ D(D+1) - 2\times (2D) = D (D-3) $$ degrees of freedom in the phase space which correspond to $\frac{1}{2}D (D-3) $ dynamical fields in the configuration space. 

According to Dirac \cite{Dirac}, the first class constraints are generators of gauge transformations which remain the total action 
 $$ S_{T}= \int dt \left( p_{i}\dot{q}_{i} -\mathcal{H}_{c}-u_{i}\phi_{i}\right) $$
 unchanged. Here $\phi_{i}$'s are primary constraints and we have considered for simplicity a system of finite number of degrees of freedom. If we wish $G= \Sigma C_{a}\phi_{a}$,  where $ \phi_{a}$'s are all first class constraints, to be the generator of gauge transformations, it should obey the following essential conditions \cite{pons}
 \begin{align}\label{pons cond}
      &\left\lbrace G, \ H_{c}\right\rbrace +\frac{\partial G}{\partial t} = PC \ , \nonumber \\
      & \left\lbrace G,\ PC \right\rbrace = PC \ , 
\end{align}
 where PC means primary constraints \footnote{These conditions are discussed in so many papers around 1990 (See for instance  \cite{Batle, Garcia, Shabani}.) In  older papers the constraints abbreviated in Eqs. (\ref{pons cond}) as PC are argued to be PFC, i.e. primary first class constraints, instead of just primary constraint. However, in a more recent paper \cite{pons}, Pons has shown that in the most general circumstances the true conditions are as shown in Eqs. (\ref{pons cond}). In our present cases for massless Fierz-Pauli in flat and de Sitter space, it turns out that the conditions (\ref{pons cond}) hold for PFC's  while for partially massless models we need to consider more general conditions as given in Eqs. (\ref{pons cond}), i.e. for PC's.}.  Conditions (\ref{pons cond}) lead to a number of certain relations among the coefficients $C_{a}$'s, which reduce the number of independent gauge functions to the number of primary first class constraints.
 
  For the current case (massless Fierz-Pauli, theory in flat space), we assume the following gauge generating functional
\begin{align}\label{gf}
  G  = \int d^{D-1}x & \ C_{00} \, \phi^{00} + C_{0i} \ \phi^{0i} +  C^{\prime}_{00} \ \bar{\chi}^{\ 00}+ C^{\prime}_{0i} \ \bar{\chi}^{\ 0i}\ .
 \end{align}
 Inserting $G $ in conditions (\ref{pons cond}) gives straightforwardly the following recursion relations
 \begin{align} \label{c1i}
   & \dfrac{\partial  C^{\prime}_{0i}}{\partial t} +  \frac{1}{2}\partial_{i} C^{\prime}_{00} +  C_{0i}=0 \ . \\ 
   &  \dfrac{\partial  C^{\prime}_{00}}{\partial t}  +  C_{00}=0 \ . 
 \end{align}
Assuming $ C^{\prime}_{00}= -2 \xi_{0}$ , $ C^{\prime}_{0i}=-\xi_{i}$ gives 
\begin{align}
G & =\int d^{D-1}x \Big(2 \partial_{0}\xi_{0} \phi^{00}+\left(  \partial_{0}\xi_{i} + \partial_{i}\xi_{0} \right)  \phi^{0i}- 2\xi_{0} \bar{\chi}^{\ 00} -  \xi_{i} \bar{\chi}^{\ 0i} \Big) \ ,
\end{align}
where $\xi_{0}$ and $\xi_{i}$  are four arbitrary gauge parameters. 
The gauge transformations of our dynamical variables can be derived directly from this generating functional as
\begin{align}
	& \delta h_{00}= \left\lbrace h_{00} , \ G\right\rbrace = \  2 \partial_{0}\xi_{0} \ ,\nonumber\\
	&  \delta h_{0i}= \left\lbrace h_{0i} , \ G\right\rbrace =  \partial_{0}\xi_{i} +\partial_{i}\xi_{0} \ ,  \nonumber \\
	& \delta h_{ij} = \left\lbrace h_{ij} , \ G\right\rbrace  = 2 \partial_{j}\xi_{i} = \partial_{j}\xi_{i}  + \partial_{i}\xi_{j}\ ,
\end{align}
which can be written in the covariant form as 
\begin{align}
\delta h_{\mu\nu} = \partial_{\mu}\xi_{\nu} +\partial_{\nu}\xi_{\mu}\ .
\end{align}
This is the well-known gauge symmetry of Fierz-Pauli theory which is, as expected, the gauge symmetry inherited from the Hilbert-Einstein theory under linearalization around the flat metric. 

Let us turn back to the massive Fierz-Pauli theory. The Poisson brackets of second level constraints $\chi^{00}$ and $\chi^{0i}$  with the primary constraints read
\begin{align}\label{chi0i}
\left\lbrace \chi^{0i} (x)  , \ \phi^{0j}(y)\right\rbrace = 2 M^{2} \delta^{ij} \delta(x-y)\ne 0 \ .
\end{align}
 Hence the set of constraints $ \phi^{0i}$ and $\chi^{0i}$ act as six second class constraints, which determine the Lagrange multipliers $u_{0i}$ in Eq. (\ref{H_{T}}). Then we should consider the constraint $\chi^{00}$ which has strongly vanishing Poisson brackets with primary constraints. Consistency condition $d\chi^{00}/d t\approx 0 $ dose not determine any Lagrange multiplier; rather, it leads to a third level constraint as
\begin{align}
\psi^{00} = -\partial_{i}\partial_{j}\pi^{ij} - \nabla^{2}\partial_{k}h_{0k} - \frac{D-3}{D-2}M^{2}\partial_{k}h_{0k} + \frac{1}{D-2} M^{2}\pi^{kk}\ .
\end{align}
Again the constraint $\psi^{00}$ has strongly vanishing Poisson brackets with the primary constraints. Similar to the previous case, consistency of $\psi^{00}$ leads to the fourth level constraint $\Lambda^{00}$ as
\begin{align}
\Lambda^{00}\equiv \left\lbrace \psi^{00}  , \  H_{c}\right\rbrace \approx \frac{1}{2}M^{4}\Big( 3 h_{00} - h_{kk}\Big)\ . 
\end{align}
Now, since 
\begin{align}\label{lamda00}
\left\lbrace \Lambda^{00}(x) , \ \phi^{00}(y)\right\rbrace = 2 M^{4} \delta(x-y)\ne 0 \ ,
\end{align}
 consistency of $\Lambda^{00}$ determines the remaining undetermined Lagrange multiplier $u_{00}$. In this way, we have found a four level chain of second class constraints as $(\phi^{00}, \ \chi^{00}, \ \psi^{00}, \ \Lambda^{00})$, besides six existing second class constraints $\phi^{0i}$ and $\chi^{0i}$.
 All of the constraint chains are self-conjugate, in terminology of ref. \cite{Loran}. It is specially interesting to see that in the first chain the Poisson bracket of the fourth (last) level constraint $\Lambda^{00}$ with the primary constraint $\phi^{00}$ is minus the Poisson bracket of the second and third level constraints $\chi^{00}$ and $\psi^{00}$, as predicted there.
 
From the formula(\ref{master equation}), the  number of dynamical phase space variables is $ D (D+1)-2(D+1)=(D+1)(D-2) $, which correspond to $ \frac{1}{2}(D+1)(D-2)$ degrees of freedom for the  massive Fierz-Pauli model.
\section{Hamiltonian structure of Fierz-Pauli action in  curved space }
In this section we investigate the Hamiltonian structure of the Fierz-Pauli theory in a background metric with constant curvature, i.e. the de Sitter or Anti de Sitter background metric.\\
Similar to flat space, the model includes, as dynamical variables, the components of a symmetric rank-2 tensor $h_{\mu\nu}$. However, the partial derivatives in Eq. (\ref{a1}) should be replaced by covariant derivatives defined according to the background metric.
 Hence, the action of the massless theory in $D$ dimension reads
 \begin{align}
  \bar{S} =  \int d^{D}x \sqrt{-g} \ \Big(  & - \frac{1}{2}\nabla_{\lambda}h^{\mu\nu}\nabla^{\lambda}h_{\mu\nu} + \nabla_{\alpha}h_{\mu\nu}\nabla^{\nu}h^{\mu\alpha} -\nabla_{\mu}h\nabla_{\nu}h^{\mu\nu} + \frac{1}{2}\nabla_{\lambda}h\nabla^{\lambda}h\nonumber\\ 
  & + \Lambda \Big(  h^{\mu\nu}h_{\mu\nu}- \frac{1}{2}h^{2} \Big)\Big) \label{ds action2} \ . 
  \end{align}
 The last term is introduced due to the nontrivial commutators of the covariant derivatives. With this modification, the action (\ref{ds action2}) is invariant under the gauge transformations
 \begin{align}
 \delta h_{\mu\nu} = \nabla_{\mu}\xi_{\nu} +  \nabla_{\nu}\xi_{\mu} \ .
 \end{align}
 Inserting mass is achieved again by a tuned Fierz-Pauli mass term as follows
 \begin{align}
 S = \bar{S} - \int d^{D}x \ \sqrt{-g} \ \frac{1}{2} M^{2} \Big( h^{\mu\nu}h_{\mu\nu}-h^{2}\Big) \ .\label{s massive}
 \end{align}
 The de Sitter metric in flat slicing coordinates is
  \begin{align}\label{de Sitter}
 ds^{2} = - d t^{2} + e^{2 m t} \sum_{i=1}^{D-1} d x_{i}^{2} \ ,
  \end{align}
  where $ m^{2} = \frac{\Lambda}{D-1} $. It is easy to find the Ricci scalar in terms of $\Lambda$ and $D$ as follow
  \begin{align}
  R = \left(\frac{2 D}{D - 2} \right) \Lambda \ .
  \end{align}
  In order to analyze the canonical structure of the model, we need to separate time and space components and expand the covariant derivatives in terms of the connections emerged from the de Sitter metric. So, the action (\ref{s massive}) may be written as follows
  \begin{align}
S = \int d^{D}x \Big(L_{2} + L_{1} + L_{0}\Big) \ , 
  \end{align}
      where $L_{2}$ and $L_{1}$ are the quadratic and linear parts of the action with respect to velocities, respectively and $L_{0}$ in independent of velocities, as follow
    \begin{align} 
 L_{2} & = \frac{1}{2} e^{(D-5)m t} (\partial_{0}h_{ij})^{2}- \frac{1}{2}e^{(D-5)m t}\left(\partial_{0}h_{kk}\right)^{2}\ ,\\
  L_{1} & = e^{-(D-5)m t}\left(\partial_{i}h_{0i}\right)\left( \partial_{0}h_{kk} \right)-2 e^{(D-5)m t}(\partial_{i}h_{0j})(\partial_{0}h_{ij})+ e^{(D-3) m t}\partial_{i}h_{00}\partial_{0}h_{0i} \nonumber \\
  & -  e^{(D-3) m t}\partial_{i}h_{0i}\partial_{0}h_{00}+ e^{(D-3)m t}h_{kk}\partial_{0}h_{00} + e^{(D-3)m t} m^{2}h_{ii}\partial_{0}h_{kk}  \nonumber \\
  & + e^{(D-3)m t} m^{2}h_{ii}\partial_{0}h_{kk}+ (3-D) e^{(D-3)m t}h_{00}\partial_{0}h_{kk} - (D-1) e^{(D-1)m t}h_{00}\partial_{0}h_{00}\ , \\ 
   L_{0} & = - \frac{1}{2} e^{(D-7)m t}(\partial_{k}h_{ij} ) ( \partial_{k}h_{ij}) + \frac{1}{2} e^{(D-7)m t}\left( \partial_{k}h_{ii} \right) \left( \partial_{k}h_{jj}\right)  + e^{(D-5)m t}(\partial_{i}h_{0j})(\partial_{i}h_{0j}) \nonumber \\
  & - e^{(D-5)m t}(\partial_{i}h_{ij})(\partial_{j}h_{00}) + e^{(D-7)m t}(\partial_{i}h_{jk})(\partial_{k}h_{ij})
   + e^{-(D-5)m t}\left( \partial_{i}h_{jj}\right)\left( \partial_{i}h_{00} \right) \nonumber \\
   & + e^{-(D-5)m t} \left( \partial_{i}h_{jj}\right)\left( \partial_{i}h_{00} \right) ‌- e^{(D-5)m t} (\partial_{i}h_{0j})(\partial_{j}h_{0i}) 
   + (D-1) e^{(D-3)m t} m^{2}h_{00}h_{kk}  \nonumber \\
   &  - M^{2} e^{(D-3)m t} h_{00}h_{kk} + (D-1)e^{(D-5)m t} m^{2}h^{2}_{ij}
   - \frac{1}{2} M^{2} e^{(D-5)m t}h^{2}_{ij}  \nonumber \\
   & - \frac{1}{2}(D-1)e^{(D-5)m t} m^{2}h^{2}_{kk}+\frac{1}{2}e^{(D-5)m t} M^{2}h^{2}_{kk}
    ‌+ 2 (1-D) e^{(D-3)m t} m^{2}h^{2}_{0k} \nonumber \\
    & + e^{(D-3)m t} M^{2} h^{2}_{0k} - \frac{1}{2}(D-1)e^{(D-5)m t} m^{2}h^{2}_{00} 
   + 4 m  e^{(D-3)m t}h_{ij}\partial_{i}h_{0j}  \nonumber \\
   & - 2 e^{(D-5)m t}h_{kk}\partial_{i}h_{0i}
   + e^{(D-5)m t}h_{0j}\partial_{i}h_{ij} 
  + (2D-6)e^{(D-3)m t} m^{2}h_{00}h_{kk} \nonumber \\
  &- 2 e^{(D-5)m t} m^{2}h^{2}_{ij}\ .
  \end{align}  
   The canonical momenta are derived as follows:
   \begin{align} 
   &\pi^{00} =- e^{(D-3) m t} \partial_{k}h_{0k} + e^{(D-3) m t} m h_{kk} + (D-1)e^{(D-1) m t} m h_{00}\ ,\nonumber \\
   & \pi^{0k}= e^{(D-5) m t} \partial_{k}h_{ii} + e^{(D-3) m t}\partial_{k}h_{00} - 4 e^{(D-3) m t} m h_{0k}\ , \nonumber \\
   & \pi^{ij} = e^{(D-5)m t}\partial_{0}h_{ij}-e^{(D-5)m t}\partial_{0}h_{kk}\delta_{ij} -2 e^{(D-5)m t}\partial_{i}h_{0j} \nonumber \\
   & \quad + e^{(D-5)m t}\partial_{k}h_{0k}\delta_{ij}+e^{(D-5)m t}m h_{kk}\delta_{ij}- \left( D-3\right) e^{(D-3)m t}m h_{00}\delta_{ij}\ .
   \end{align}
      As is seen, the primary constraints for the massless, as well as massive, theory read
      \begin{align}\label{consts} 
      &\phi^{00} = \pi^{00}+ e^{(D-3) m t} \partial_{k}h_{0k} -e^{(D-3) m t} m h_{kk} - (D-1)e^{(D-1) m t} m h_{00} \ , \nonumber \\
      & \phi^{0i}=\pi^{0i}- e^{(D-5) m t} \partial_{i}h_{kk} -e^{(D-3) m t}\partial_{i}h_{00} + 4 e^{(D-3) m t} m h_{0i} \ .
      \end{align}
       The canonical Hamiltonian for the massless theory is derived as follows 
    \begin{align}\label{Hamiltonian}  
   \bar{H}_{c}  =\int d^{D}x \Big[ & \frac{1}{2} e^{-(D-5) m t}(\pi^{ij})^{2} - \frac{1}{2}\frac{1}{D-2}e^{-(D-5)m t} (\pi^{kk})^{2} +  \pi^{ij}\left(  \partial_{i}h_{0j}+ \partial_{j}h_{0i} \right) + 2 e^{(D-5) m t}  m^{2} h^{2}_{ij}  \nonumber \\
     & + \frac{1}{D-2} m \pi^{kk} h_{ii}- \frac{1}{D-2}\pi^{kk}\partial_{i}h_{0i} + 2 e^{(D-5) m t}  m^{2} h^{2}_{ij}- 2 \left( D-3 \right)  e^{(D-3) m t} m^{2} h^{2}_{0k}    \nonumber \\
      & + \frac{1}{2} e^{(D-7) m t}(\partial_{k}h_{ij} ) ( \partial_{k}h_{ij}) - \frac{1}{2} e^{(D-7) m t}\left( \partial_{k}h_{ii} \right) \left( \partial_{k}h_{jj}\right)+ e^{(D-7) m  t}\left( \partial_{i}h_{ij}\right)\left( \partial_{j}h_{kk} \right) \nonumber \\
                      & -e^{(D-7) m t}\left( \partial_{i}h_{jk}\right)\left( \partial_{k}h_{ij} \right)   - \frac{D-1}{D-2}e^{(D-5) m t}\left( \partial_{k}h_{0k}\right)^{2} - 4 e^{(D-5) m t} m\left( \partial_{i}h_{0j}\right)h_{ij} \nonumber \\
                   & + \frac{D^{2}-4D+5}{D-2}e^{(D-5) m t} m h_{ii}(\partial_{k}h_{0k}) - \frac{1}{2} \frac{D-1}{D-2}e^{(D-5) m t} m^{2} h^{2}_{kk}  \nonumber \\
                        & -h_{00}\Big(   \frac{\left(D-1 \right) \left( D^{2}-5D+7\right) }{2\left(D-2 \right) } m^{2} e^{(D-1) m t}h_{00} - \frac{2D^{2}-9D+11}{D-2} e^{(D-3)m t} m (\partial_{i} h_{0i})               
                                       \nonumber \\
                                   &       - \frac{D-3}{D-2} m e^{2 m t} \pi^{kk}  +  e^{(D-5) m t}\left( \nabla_{i}h_{kk}\right)-e^{(D-5)m t}\left( \partial_{i}\partial_{j}h_{ij}\right) \nonumber \\
                                  &   - \frac{2D^{2}-9D+11}{D-2} e^{(D-3)m t} m^{2} h_{kk} \Big) +  e^{(D-5)m t} \left( \partial_{i}h_{0j}\partial_{i}h_{0j}+\partial_{i}h_{0j}\partial_{j}h_{0i}\right) \Big] \ , 
     \end{align}
While for massive model the canonical Hamiltonian is 
\begin{align}
H_{c}=  \bar{H}_{c}+ \int d^{D}x \left[ M^{2}e^{(D-5) m t}\Big( e^{2 m t} h_{kk} h_{00} + \frac{1}{2} h^{2}_{ij} - \frac{1}{2}  h^{2}_{kk}- e^{2 m t}  h^{2}_{0k} \Big) \right] \ .
\end{align}
Now let us proceed each case one by one.
\subsection{Massless model}
As usual we should investigate the consistency of the primary constraints. However, an interesting and noticeable point is explicit time dependence of the constraints, which implies different formula for their consistency as follows   
   \begin{align}\label{a5}
   &\bar{\chi}^{\ 00} =  \left\lbrace  \phi^{00}  ,\  \bar{H}_{T} \right\rbrace  + \dfrac{\partial\phi^{00}}{\partial t} \ ,\nonumber\\
   & \bar{\chi}^{\ 0i} =  \left\lbrace  \phi^{0i}  ,\  \bar{H}_{T} \right\rbrace  + \dfrac{\partial\phi^{0i}}{\partial t} \ .
   \end{align} 
   The  total Hamiltonian is completely similar to $H_{T}$ of Eq. (\ref{H_{T}}) where $\phi^{00}$ and $\phi^{0i}$ should be inserted from Eqs. (\ref{consts}).
   Since the primary constraints commute, the total Hamiltonian $\bar{H}_{T}$  should be replaced by $\bar{H}_{C}$ in Eq (\ref{Hamiltonian})and the final result for the second level constraints read
      \begin{align} 
                    \bar{\chi}^{\ 00} & = m e^{2 m t}\pi^{kk} - e^{(D-5)m t}\partial_{i}\partial_{j}h_{ij} + (D-1) e^{(D-3)m t} m \partial_{k}h_{0k}+ e^{(D-5)m t}\nabla^{2}h_{kk}  \nonumber \\
                    & + e^{(D-3)m t}(D-3)m^{2}h_{kk}- (D-1) e^{(D-1) m t}m^{2}h_{00} \ , \nonumber\\
                     \bar{\chi}^{\ 0i} & = 2 \partial_{j}\pi^{ij} +2 e^{(D-5)m t}\partial_{i}h_{kk} - 2  e^{(D-3)m t} m \partial_{i}h_{00} - 4 m e^{(D-5)m t}\partial_{j}h_{ij} \nonumber \\
                     & + 2 e^{(D-5) m t}\nabla^{2} h_{0i} \ .                                      
                       \end{align} 
We can directly see that the second level constraints have vanishing Poisson brackets with the primary ones. Moreover, they have weakly vanishing time derivatives as follows
   \begin{align}
&\dfrac{d\bar{\chi}^{\ 00}}{d t} =\left\lbrace  \bar{\chi}^{\ 00}  ,\  \bar{H}_{T} \right\rbrace  + \dfrac{\partial \bar{\chi}^{\ 00}}{\partial t}= -\dfrac{1}{2}\partial_{i} \bar{\chi}^{\ 0i} \approx 0 \ , \nonumber \\
& \dfrac{d\bar{\chi}^{\ 0i}}{d t}=\left\lbrace  \bar{\chi}^{\ 0i}  ,\  \bar{H}_{T} \right\rbrace  + \dfrac{\partial \bar{\chi}^{\ 0i}}{\partial t} = -2 m \bar{\chi}^{\ 0i}\approx 0 \ .
 \end{align}
     So, similar to flat space, we have $ 2 D $ first class constraints and $$\frac{1}{2}\left( D(D+1) - 2 \times 2 D\right)   = \frac{1}{2} D(D-3)$$ dynamical degrees of freedom.\\
     Assuming the gauge generating functional similar to Eq. (\ref{gf}) and inserting it in conditions (\ref{pons cond}) gives the following relations among the coeficients
 \begin{align} \label{c1i} & \dfrac{\partial C^{\prime} _{0i}}{\partial t}  + \frac{1}{2}\partial_{i}  C^{\prime}_{00} - 2 m C^{\prime}_{0i}+  C_{0i}=0 \ , \\  &  \dfrac{\partial  C^{\prime}_{00}}{\partial t}  +  C_{00}=0 \ . 
\end{align} 
Hence the gauge generating functional reads
 \begin{align} \label{Gds }
         G  = \int d^{D-1}x  \Big(  2\partial_{0}\xi_{0}\phi^{00}
          + \left(  \partial_{0}\xi_{i}+ \partial_{i}\xi_{0}-2m \xi_{i} \right) \phi^{0i}
         - 2 \xi_{0}\chi^{00} - \xi_{i}\chi^{0i}\Big)\ , 
 \end{align} 
 The gauge transformations of our dynamical variables in curved space can be derived from  generating functional (\ref{Gds }) as follows
         \begin{align}
                  &\delta h_{00} = \left\lbrace h_{00} , \ G\right\rbrace  =   2 \partial_{0}\xi_{0} = 2\nabla_{0}\xi_{0}\ , \nonumber \\
                           & \delta h_{0i}= \left\lbrace h_{0i} , \ G\right\rbrace  =   \left( \partial_{0}\xi_{i} + \partial_{i}\xi_{0}- 2 m \xi_{i} \right)= \nabla_{0}\xi_{i} + \nabla_{i}\xi_{0}  \ ,  \nonumber \\
                          & \delta h_{ij} = \left\lbrace h_{ij} , \ G\right\rbrace   = - 2 m e^{2 m t} \xi_{0} \delta_{ij} + 2 \partial_{j}\xi_{i} =  \nabla_{i}\xi_{j}+ \nabla_{j}\xi_{i} \ .
                  \end{align}
                  \subsection{Massive model}
          Now let us consider the massive theory. In this case the second level constraints change to
     \begin{align}
     \chi^{00} = \bar{\chi}^{ \ 00}- M^{2} e^{(D-3)m t}h_{kk} \ , \quad
          \chi^{0i} =  \bar{\chi}^{ \ 0i} + 2 M^{2}e^{(D-3)m t}h_{0i} \ .
     \end{align}
     Emerging the mass term in the constraints $\chi^{0i}$ makes them of second class with the primary constraints $\phi^{0i}$ as
     \begin{align}
     \left\lbrace \chi^{0i}  , \ \phi^{0j}\right\rbrace = 2 M^{2} e^{(D-3)m t}\delta^{ij}\ .
     \end{align}
     This leads to determining the Lagrange multipliers $u_{0i}$ in total Hamiltonian. However, $\chi^{00}$ still has vanishing Poisson brackets with the existing constraints.
     Consistency of $\chi^{00}$ gives
     \begin{align}
    \psi^{00}  = & -\partial_{i}\partial_{j}\pi^{ij} + \frac{1}{D-2}e^{2 m t}M^{2}\pi^{kk} + 2 e^{(D-5)m t}m \partial_{i}\partial_{j}h_{ij} - \frac{D-3}{D-2}e^{(D-3)m t} M^{2} \partial_{k}h_{0k}\nonumber \\
    & - \frac{1}{D-2} e^{(D-3)m t} m M^{2} h_{kk} - \frac{D-1}{D-2} e^{(D-1)m t}m M^{2} h_{00} - e^{(D-5)m t} \nabla^{2}\partial_{k}h_{0k} \nonumber \\
    & - e^{(D-5) m t} \nabla^{2}h_{kk} + e^{(D-3) m t} m \nabla^{2}h_{00}
     \end{align}
     as the third level constraint.
     
There is an important subtlety at this point which needs more care.  Second and third level constraints  $\chi^{00}$ and $\psi^{00}$ have non-vanishing Poisson brackets with the emerging constraints, so  it seems that the consistency procedure stops here. However, this is not a correct conclusion.

 In fact, the primary constraints $\phi^{0i}$ have been categorized previously as second class constraints together with the second level constraints $\chi^{0i}$. So it is not legitimate to collect them again as conjugate second class constraints with some other constraint. In other words, if a function, which is conjugate to some constraint, appears several times in different constraints, all of them would have non-vanishing Poisson brackets with the corresponding constraint. Hence, it is necessary to remove such a function from the emerged constraints by redefining them, i.e. adding suitable combinations of the previous constraints to them.
 
 For the case at hand, we should redefine the second level constraint as
     \begin{align}\label{a6}
   {\chi}^{\prime \, 00}= \chi^{00} + \partial_{i}\phi^{0i} \ ,
     \end{align}
     and the third level constraint as
     \begin{align}\label{a7}
        {\psi}^{\prime\prime \, 00}= \psi^{\prime \, 00} + m\partial_{i}\phi^{0i} \ ,
          \end{align}
          where $$ \psi^{\prime \, 00}=\psi^{00} + \partial_{i}\chi^{0i} \ .$$ 
These redefinitions lead to the following constraint algebra 
\begin{align} \label{a8}
        &\left\lbrace \phi^{00} , \ \phi^{0i}  \right\rbrace =0 \ ,  \left\lbrace \phi^{00} , \ \chi^{0i}  \right\rbrace =0 \ ,  \nonumber \\
            & \left\lbrace \chi^{\prime 00} , \ \phi^{00}  \right\rbrace =0 \ ,  \left\lbrace \chi^{\prime 00} , \ \phi^{0i}  \right\rbrace =0 \ ,  \left\lbrace \chi^{\prime 00} , \ \chi^{0i}  \right\rbrace =0 \ , \nonumber \\
                & \left\lbrace \psi^{\prime\prime \, 00} , \ \phi^{00}  \right\rbrace =0 \ ,\left\lbrace \psi^{\prime\prime \, 00} , \ \chi^{0i}  \right\rbrace =0 \ , \left\lbrace \psi^{\prime\prime \, 00} , \ \chi^{0i}  \right\rbrace =0 \ , 
         \left\lbrace \psi^{\prime\prime \, 00} , \ \phi^{0i}  \right\rbrace =0 \ ,   \nonumber \\
             &\left\lbrace \psi^{\prime\prime \, 00} , \ \chi^{\prime 00}  \right\rbrace =\frac{D-1}{D-2}M^{2}\left(M^{2}-(D-2) m^{2} \right) e^{(D-3)m t} .
\end{align}
Assuming $M^{2}\ne 0$ and $M^{2} \ne (D-2) m^{2}$ , we can  continue with consistency of ${\psi}^{\prime\prime \, 00}$. This gives the fourth level constraint $\Lambda^{00}$ as follows
\begin{align} \label{Lambda}
     \Lambda^{00} = \frac{(D-1)}{D-2}M^{2} \left(  \frac{1}{D-1} \chi^{00}+ e^{(D-3)m t}\left(   h_{kk}- e^{2 m t} h_{00} \right)\left( M^{2}-(D-2) m^{2}\right)\right) \ .
\end{align}      
This is the end of the story for the massive Fierz-Pauli theory in the de Sitter background. The problem is more or less similar to Fierz-Pauli theory in flat space, i.e. we have three constraint chains with two levels $ \left(\phi^{0i}, \ \chi^{0i} \right) $  and  one constraint chain with four levels, i.e. the constraints $ (\phi^{00}, \ \chi^{\prime \ 00}, \ {\psi}^{\prime\prime \ 00}, \ \Lambda^{00}) $. This leads to $ \frac{1}{2} \left( D(D+1)-(2 D + 2)\right)  = \frac{1}{2}(D+1)(D-2) $ dynamical degrees of freedom, as expected.\\
\subsection{Partially massless model}
Our detailed Hamiltonian analysis for the massive Fierz-Pauli theory in de Sitter background enables us to understand better the mechanism of appearing a new gauge symmetry in the special case of partially massless theories. After our novel redefinition (\ref{a6}) and (\ref{a7}) the explicit form of the last constraint $\Lambda^{00}$ in Eq. (\ref{Lambda}) brings the possibility of (weakly) vanishing  $ \Lambda^{00} $ when the Higuchi bound is saturated, i.e.
      $ M^{2} = (D-2) m^{2} = \left( \frac{D-2}{D-1}\right) \Lambda $\ . If this is the case, the consistency condition terminates at the third level and does not go further to fourth level. In fact, vanishing  of $\Lambda^{00}$ (on the boundary $ M^{2}= (D-2) m^{2} $) means that consistency of ${\psi}^{\prime\prime \,  00}$ neither determines a Lagrange multiplier nor introduces a fourth level constraint. Hence, we have one first class chain of constraints in three levels as $ (\phi^{00}, \ \chi^{\prime \, 00}, \ {\psi}^{\prime\prime \, 00}) $. Using the  formula (\ref{master equation}) with  three first class and six second class constraints gives  $$\frac{1}{2}(D(D+1) - 2 \times 3 - 2(D-1))  =\frac{1}{2}(D+1)(D-2) -1 $$ degrees of freedom.
      
     This is the famous result for partially masslessness. In fact, if someone has had done the canonical analysis of this paper, the idea of partially masslessness would be emerged with much less intelligence used by Higuchi in his pioneer paper\cite{Higuchi}.
     
     Having a chain of first class constraints at hand, enables us to write the generator of gauge transformation which they generate.
     Assume the gauge generating functional as
     \begin{align}
 G  = \int d^{(D-1)}x \ & C_{00} \ \phi^{00} +C^{\prime}_{\ 00} \ \chi^{\prime \ 00} +C^{\prime \prime}_{\ 00} \ \psi^{\prime\prime \ 00} \ .\label{pm-generator1}
     \end{align}      
     Inserting $G$ in conditions (\ref{pons cond})  gives the following recursion relations among the coefficients $C_{00}$, $C^{\prime}_{\ 00}$ and $C^{\prime \prime}_{\ 00}$.
     \begin{align} \label{C pm1}
   & \dfrac{\partial  C^{\prime}_{00}}{\partial t} +  \frac{1}{D-2} M^{2} C^{\prime\prime}_{00} +  C_{00}=0 \ , \\ \label{C pm2}
   &  \dfrac{\partial  C^{\prime\prime}_{00}}{\partial t}  +  C^{\prime}_{00}=0 \ . 
 \end{align}
     Considering  $C^{\prime \prime}_{\ 00} = \alpha(x)$ as an arbitrary gauge parameter, Eqs. (\ref{C pm1}) and (\ref{C pm2}) imply
     \begin{align}
                C^{\prime}_{\ 00}= -\partial_{0}\alpha(x)\ ,  \ C_{00}= \left( \partial_{0}\partial_{0} - \dfrac{1}{D-2}M^{2}\right)\alpha(x)\ .
                \end{align}
   Therefore Eq. (\ref{pm-generator1}) reads
               \begin{align}
     G  = \int d^{D-1}x \   \phi^{00} \Big( \partial_{0}^{2} - \frac{1}{(D-2)} M^{2}\Big) \alpha(x) 
      - \chi^{\prime \ 00} \  \partial_{0}\alpha(x)  + \psi^{\prime\prime \ 00} \alpha(x) \ .  \label{pm-generator2}
     \end{align}
         Using this functional, it is easy to find the gauge transformations of the field components as follows
     \begin{align}
     &\delta h_{00}= \left\lbrace h_{00} , \ G\right\rbrace  = \left(  \nabla_{0}\nabla_{0} - \frac{1}{D-2} M^{2}\right)  \alpha(x)\ , \nonumber\\
     &  \delta h_{0i}= \left\lbrace h_{0i} , \ G\right\rbrace  =   \nabla_{0}\nabla_{i}\alpha(x)\ ,\nonumber\\
     &\delta h_{ij} = \left\lbrace h_{ij} , \ G\right\rbrace  = \left(  \nabla_{i}\nabla_{j} + \frac{1}{D-2}M^{2} e^{2 m t}\delta_{ij}\right)   \alpha(x)\ ,
     \end{align}
     which can be written covariantly as
     \begin{align}
     \delta h_{\mu\nu} = \Big( \nabla_{\mu}\nabla_{\nu} +  \frac{1}{D-2} M^{2}g_{\mu\nu}\Big) \alpha\ .
     \end{align}
     \section{Conclusions}  
     In this paper we tried to give the Hamiltonian analysis of Fierz-Pauli theory in flat and curved spaces in full details. The main aim of this analysis goes beyond counting the number of degrees of freedom. In fact, the underlying algebraic behavior of the constraints helps us to better understand the mechanism of a gauge symmetry. For example the remarkable difference between the massive and massless Fierz-Pauli theories, in the number of degrees of freedom, as well as the gauge symmetry,  lies exactly on the presence of the mass term on the right hand side of the Poisson brackets of the second level constraints (Compare Eq.(\ref{massless case}) with Eqs. (\ref{chi0i}) and (\ref{Lambda})).  Hence, in our opinion the constraint analysis provides a powerful tool for observing step by step what happens.
     
     For the dynamical behavior of the system in different regions of the parameter space, one interesting class of models which have different dynamical behavior for special value of the parameters of the model are PM theories. These are massive Fierz-Pauli spin-2 fields propagating in a curved space with constant curvature. We showed what happens to the constraint algebra of the theory exactly when the mass of graviton matched to the cosmological constant. As we observed in Eqs. (\ref{a8}) and Eq. (\ref{Lambda}), when $M^{2}=\left(  D-2 \right) m^{2}$, one constraint chain changes its characteristics, i.e. a four-element second class chain  transmutes to a three-element first class chain. 
     
     Another noticeable advantage of the Hamiltonian analysis is to find the explicit relationship between the first class constraints and the gauge symmetries of the system. This can be achieved by introducing the generating functional of gauge transformations via a definite combination of first class constraints and the gauge parameters. For this reason we followed the procedure given in Refs.\cite{Shabani,pons}  
     both for Fierz-Pauli models (in flat and curved space) and for PM model.
     
     Some technical methods employed in this paper are noticeable. One of them is treating the explicit time dependent constraints, which needs to consider the appropriate  derivative terms. Another point is finding a suitable redefinition of a constraint by adding appropriate combination of the constraints derived already. See Eqs. (\ref{a6}) and (\ref{a7}) for introducing  equivalent constraints $\chi^{\prime \, 00}$ instead of $\chi^{00}$ and $\psi^{\prime\prime \, 00}$ instead of the emerged one $\psi^{00}$.
     Although these techniques seem to be simple when one skims over them, however, in practice these subtle points prevent people to employ the Hamiltonian analysis.\\
     
      \section{Acknowledgment}  
      The authors thank Prof. S Deser for his valuable comments.

     \end{document}